\documentclass[fleqn,usenatbib]{mnras}
\usepackage{newtxtext,newtxmath}
\usepackage[T1]{fontenc}
\usepackage{url}
\DeclareRobustCommand{\VAN}[3]{#2}
\let\VANthebibliography\thebibliography
\def\thebibliography{\DeclareRobustCommand{\VAN}[3]{##3}\VANthebibliography}
\usepackage{graphicx}	
\usepackage{amsmath}	
\title[Starburst Galaxy TeV Emission]
{Could the TeV emission of starburst galaxies originate from pulsar wind nebulae?}

\author[Xiao-Bin Chen et al.]{
Xiao-Bin Chen,$^{1,2}$
Ruo-Yu Liu,$^{1,2}$\thanks{Email: ryliu@nju.edu.cn}, Xiang-Yu Wang,$^{1,2}$ and Xiao-Chuan Chang$^3$
\\
$^{1}$School of Astronomy and Space Science, Nanjing University, Nanjing 210023, People's Republic of China;\\
$^2$Key Laboratory of Modern Astronomy and Astrophysics, Nanjing University, Ministry of Education, Nanjing 210023, People's Republic of China;\\
$^3$ School of Physics, Xi'an Jiaotong University, Xi'an 710049, People's Republic of China
}

\date{Accepted XXX. Received YYY; in original form ZZZ}
\pubyear{2023}

\begin{document}
\label{firstpage}
\pagerange{\pageref{firstpage}--\pageref{lastpage}}
\maketitle

\begin{abstract}
While the GeV $\gamma$-ray emission of starburst galaxies (SBG) is  commonly thought to arise from hadronic interactions between accelerated cosmic rays and interstellar gas, the origin of the TeV $\gamma$-ray emission is more uncertain. One possibility is that a population of  pulsar wind nebulae (PWNe) in these galaxies could  be responsible for the TeV $\gamma$-ray emission. 
In this work, we first  synthesize a PWNe population in the Milky Way, and assessed their contribution to the $\gamma$-ray emission of the Galaxy, using a time-dependent model to calculate the evolution of the PWN population.
Such synthetic PWN population can reproduce the flux distribution of identified PWNe  in the Milky Way given a distribution of the initial state of the pulsar population. 
We then apply it to starburst galaxies and quantitatively calculate the spectral energy distribution of all PWNe in the SBG NGC 253 and M82. We propose that TeV $\gamma$-ray emission in starburst galaxies can be dominated by  PWNe for a wide range of parameter space. The energetic argument requires that $\eta_e \times v_{\rm SN} > 0.01 {\rm yr}^{-1}$, where $\eta_e$ is the fraction  the spin-down energy going to electrons and  $v_{\rm SN}$ is the supernova rate. By requiring the synchrotron emission flux of all PWNe in the galaxy not exceeding the hard X-ray measurement of NGC 253, we constrain the initial magnetic field strength of PWNe to be $\la 400 \mu$G. 
Future observations at higher energies with LHAASO or next-generation neutrino observatory IceCube-Gen2 will help us to understand better the origin of the TeV $\gamma$-ray emission in SBGs.
\end{abstract}

\begin{keywords}
galaxies: starburst -- gamma-rays: galaxies -- pulsars: general 
\end{keywords}



\section{Introduction}

TeV $\gamma$-ray have been detected from the starburst galaxies NGC 253 \citep{2012ApJ...757..158A} and M82 \citep{2009Natur.462..770V}. Starburst galaxies have a very high star-formation rate (SFR) in a compact central region, the starburst nucleus (SBN). 
SBGs are endowed with a high interstellar medium (ISM) gas density ($n_\text{ISM} \sim 100 - 1000~ \text{cm}^{-3}$), and magnetic fields of the order of $100~\mu$G \citep{2006ApJ...645..186T}.
The enhanced star-forming activity leads to a high supernova (SN) rate $v_\text{SN} \sim 0.1 - 1 ~\text{yr}^{-1}$, and subsequently a high injection luminosity of cosmic rays (CRs) including hadrons and leptons. CRs could produce $\gamma$-ray from GeV to TeV by interacting with the local ISM and radiation fields (soft photons) \citep{2008A&A...486..143P}.

Many studies \citep{2019MNRAS.487..168P, 2018MNRAS.474.4073W, 2020MNRAS.493.2817K, 2021ApJ...907...26H} recently modelled the CR spectrum in SBGs based on their multi-wavelength spectra. They found that electrons typically effectively lose energy within SBN, while it is uncertain whether SBNs are calorimeters for protons. It depends on the properties of CR transport in the ISM such as the diffusion coefficient and the speed of the galactic wind driven from the SBN. 
For example, \cite{2019MNRAS.487..168P} consider the SBN as a spherical compact region with a leaky-box model, in which the injection of CRs is balanced by energy losses, advection, and diffusion. Under several assumptions and reasonable values for the parameters, it is found that the hadronic model can reproduce the radiation spectrum of SBG NGC253 and M82 from  GeV to a few TeV. 

On the other hand, the TeV $\gamma$-ray survey of the Galactic plane reveals multiple classes of sources contributing to the total $\gamma$-ray luminosity. This poses a question that how is the fraction of their contribution to the total $\gamma$-ray flux observed from SBGs. Among various TeV $\gamma$-ray emitters in our Galaxy, PWNe are the dominant class.
The energies of pulsar wind electrons and positrons range from $\sim$ 1 GeV to $\sim$ 1 PeV, placing their synchrotron and inverse Compton (IC) emission into radio-X-ray and GeV-TeV bands, respectively. 
According to \citet{2010AIPC.1248...25K}, about 60 PWNe associated with known radio or $\gamma$-ray pulsars
have been detected, 33 of which have measured TeV fluxes.
The birthrate of pulsars is correlated with the star formation rate, and thus PWNe should be more abundant in starbursts galaxies. 
Based on this, \citet{2012APh....35..797M} argued that a population of individual PWNe could be responsible for the detected TeV emission from SBGs. They used the properties of Galactic TeV-detected PWNe to estimate the contribution of leptonic $\gamma$-ray emission. Considering the impact of starburst environment on PWNe,
\citet{2013MNRAS.429L..70O} revisited this problem and suggested that PWNe can make a significant contribution to the TeV fluxes, provided that the injection spectrum of particles is sufficiently hard and that the average pulsar birth period is  short ($\sim$ 35 ms). 
However, these works assumed that all PWNe are the same, without considering the difference among individual PWNe and the possible parameter distribution among them.

In this work, we revisit the $\gamma$-ray emission contributed by the PWN population in SBGs, considering the time-dependent evolution of the PWN population in SBGs.
We will also discuss the influence of model parameters that are not well determined  on the TeV emission of SBGs. The rest of the paper is organized as follows: 
in Section~\ref{sec:PWNe}, we describe and discuss the assumptions for the PWNe population model; 
in Section~\ref{sec:MW}, we use the model in Milky Way and compare with observations available;
in Section~\ref{sec:apply}, we apply the model to the starburst region and place constraints on the parameters;
Then, we give discussions in Section~\ref{discussion} and conclusions in Section~\ref{conclusions}. We note that the work in this paper appears in earlier form in the conference report by \citet{Chen:2023uK}.

\section{The PWN model} \label{sec:PWNe}

The time-dependent evolution of relativistic electrons and positrons pairs
\footnote{Hereafter we do not distinguish positrons from electrons for simplicity.}
in PWNe can provide us details about the radiation spectrum as a function of their ages. Cooling of these electrons in the magnetic and radiation fields leads to a multi-wavelength spectrum from radio to $\gamma$-ray.
\citet{2018A&A...612A...2H} introduce a time-dependent model for PWNe. It allows us to trace the evolution of the very-high-energy (VHE; $E>100\,$GeV) electron population, and hence the radiative output of a PWNe, based on a few general assumptions. We adopt this PWN model, then extend it to the evolution of PWN in SBG. Its essential traits are outlined in the following.

PWNe are powered by the rotational energy of related pulsars, which spin down with time and convert a fraction of their rotational energy into non-thermal particles. Over time the pulsar's rotation slows down, and the energy input rate into the nebula decreases. The spin-down luminosity of the pulsar is the rate at which rotational kinetic energy is dissipated, and is thus given by the equation:
\begin{equation}
    \dot{E}(t) \equiv I\Omega\dot{\Omega}=4\pi^2 I \dot{P}(t)/{P^3(t)}
\end{equation}
where $\Omega$ is the angular velocity of the pulsar's spin and $I$ is the neutron star's moment of inertia. For a typical mass of $1.4~M_{\sun}$  and a radius of 10~km, we have $I\sim 10^{45}~\text{g}~\text{cm}^{-2}$.
The braking index is assumed to be $n = 3$ for all pulsars in this work, corresponding to a spin-down due to the dipole radiation only. Then pulsar's the surface magnetic field at the equatorial plane $B_{\rm s}$ can be calculated through the observed spin periods \textit{P} and the time derivatives $\dot{P}$ by \citet{2006ARA&A..44...17G}
\begin{equation} \label{equation_B}
    B_{\rm s}\equiv3.2\times10^{19}(P\dot{P}/{\rm s})^{1/2}\text{G}.
\end{equation}

For the energy spectrum of electrons freshly injected into the nebula we assume the following power-law shape:
\begin{equation}
    \frac{\text{d} N_{\text{inj}}}{\text{d} E}(E, t)=\Phi_0(t)\left(\frac{E}{1 \text{TeV}}\right)^{-\alpha}
\end{equation}
with a power-law index $\alpha$. $\Phi_0(t)$ can be calculated imposing
\begin{equation}
    \int_{E_{\text{min}}}^{E_{\text{max}}}  \frac{\text{d}N_{\text{inj}}}{\text{d}E}(E,t)\text{d}E = \eta_e \dot{E}(t)
\end{equation}
where $\eta_e$ is the electron conversion efficiency, i.e. the fraction of the spin-down energy going to electrons.

The energy distribution of the electrons is assumed to be a power-law in the energy range from $E_\text{min}$ to $E_\text{max}$.
Varying the boundary energies essentially changes the number of particles contained in the inverse Compton (IC) relevant energy range, therefore they are also important parameters that will be discussed in this work. 
A low-energy break in the injection spectrum is sometimes assumed \citep[e.g.][]{2014JHEAp...1...31T}, but it only impacts the lower ends of the emission spectra. We omit it here because it does not influence the TeV flux.

Generally, the magnetic field in the PWNe is a function of both the distance \textit{r} from the pulsar and the evolution time of the PWNe. If one only cares about the total emission from a PWN, one can simply consider the average magnetic field in the PWNe which changes only with  the age of the PWN. Therefore, following \citet{2008ApJ...676.1210Z}, the magnetic field evolution is given by
\begin{equation}
    B(t) = \frac {B_0} {1 + (t/\tau_0)^{\delta_B}},  
\end{equation}
where $B_0$ is initial magnetic field of PWNe, $\delta_B$ is the B-field parameter and $\tau_0 = P_0/2\dot{P_0}$ is the characteristic timescale, adopting an index of $\delta_B = 0.6$ which satisfies the conservation of magnetic flux.

As far as energy losses are concerned, we include the possibility that particles can leave the nebula as a result of diffusion, as well as the synchrotron  energy loss and IC scattering energy loss.
The cooling of the electrons during a time step $\delta t$ is implemented in the model by means of an exponential function:
\begin{equation}
    \frac{\text{d}N_\text{cooled}}{\text{d}E}(E,t) = 
    \frac{\text{d}N}{\text{d}E}(E,t-\delta t)\cdot
    \exp\left(-\frac{\delta t}{\tau_\text{eff}(E,t)}\right).
\end{equation}
This approach uses an effective cooling timescale $\tau_\text{eff}$, which is adopted from Appendix of \citet{2018A&A...612A...2H}. 


The framework allows us to calculate the energy distribution of the electrons contained in the PWN at any given time. More specifically, the number of leptons with energy $E$ residing in the nebula at a time $t+\delta t$ is determined by the balance of freshly injected leptons and those cooled out of the respective energy interval,
\begin{equation} \label{eq:E_t}
    \frac{\text{d}N}{\text{d}E}(E,t+\delta t)=
    \frac{\text{d}N_\text{cooled}}{\text{d}E}(E,t)+
    \frac{\text{d}N_{\text{inj}}}{\text{d}E}(E,t+\delta t).
\end{equation}
The iterative evaluation of Eq.~\ref{eq:E_t} then yields the lepton energy distribution as a function of time. After considering the time-dependent evolution of relativistic particles, electrons in PWNe produce synchrotron and IC emission from radio to VHE $\gamma$-ray energies through interactions with magnetic and radiation fields, respectively.
The physics of these processes is described in the comprehensive review article by \citet{1970RvMP...42..237B}, which we follow in the implementation of the radiation mechanisms in our model.

\section{Modeling the PWN population in the Milky Way} \label{sec:MW}

Modeling of the PWN population starts with the random generation of a pulsar population.
For each pulsar, an initial pulsar spin period $P_0$ is sampled from a normal distribution with average value $\mu_{P_0}=50$~ms and the standard deviation $\sigma_{P_0}=35$~ms, truncated at 10~ms. 
These distributions are similar to the ones used by both  \citet{2011ApJ...727..123W} and \citet{2020MNRAS.497.1957J}. 
We use a typical surface magnetic field strength of pulsar at birth, i.e. $\mu_{\rm log(\textit{B}_s/G)}=12.65$
as an average value  with a log-normal distribution of the standard deviation $\sigma_{\rm log(\textit{B}_s/G)} = 0.55$ with $B$ in Gauss (similar to the values used in \citealt{2022A&A...666A...7M, Fiori_2022, 2011ApJ...727..123W}).
Assuming typical values of $10^{45}~\text{g}~\text{cm}^{-2}$ and 12~km for the neutron star inertia and radius, these properties determine the spin-down history of each pulsar. They set the maximum power available at each time for the non-thermal particle injection into PWNe.

We apply the PWN population model to the Galaxy and simulate the expected PWNe TeV flux distribution. 
The expected generation rate of pulsars in the Galaxy is $\sim$ 1/100 yr \citep{2006ApJ...643..332F}.
We can roughly estimate that around $\sim 1000$ $\gamma$-ray-emitting PWNe are generated over a period of $t_\text{end}=10^5~\text{yr}$.

By varying the final age, we have verified that $t_\text{end}$ does not affect the results of the present study, given that the collective luminosity of older PWNe is sub-dominant.

\subsection{The spatial distribution} \label{sub:distribution}
We are dealing here with only the young pulsars with ages within 100~kyr. We simply assume that pulsars are born in the Galactic plane and do not move significantly over the first $\sim$ 100~kyr of their lifetime, so we can ignore the pulsar's proper motion \citep{2002ApJ...568..289A, 2005MNRAS.360..974H}. The Galactic radial distribution of pulsars is given by \citet{2020MNRAS.497.1957J}
\begin{equation}
    \rho_r(R)=K_rR^i\text{e}^{-R/\sigma_r},
\end{equation}
where $\rho_r(R)$ is the density of pulsars (per kpc$^2$) at radius \textit{R} (in kpc) from the Galactic Centre and $K_r$, \textit{i}, and $\sigma_r$ are constants with values of 64.6~kpc$^{-2}$, 2.35, and 1.258~kpc, respectively.
We give the spatial coordinates of each PWN by this radial distribution function, and randomly assign the angular coordinates. We ignore the spiral arm structure of the Milky Way, which will not affect our results.

\subsection{The TeV flux distribution} \label{sbusec:gamma}
According to the electron energy distribution in the Section \ref{sec:PWNe}, the emission arising from synchrotron emission and IC scatterings, which are  the most important processes, can be obtained. 
The target photon fields  for IC scattering include cosmic microwave background (CMB), starlight, and infrared photons. The uniform CMB component is modelled as a black-body spectrum with an energy density of 0.26\,eV~cm$^{-3}$ and temperature of 2.7\,K. 
The starlight and infrared components can be adopted from the GALPROP code by \citet{2005ICRC....4...77P}.
The energy densities of the starlight and infrared fields are 1.92\,eV~cm$^{-3}$ and 1.19\,eV~cm$^{-3}$, respectively. The temperatures at the spectral peaks are 7906\,K for the starlight field component and 107\,K for the infrared.

\begin{figure}
    \includegraphics[width=\columnwidth]{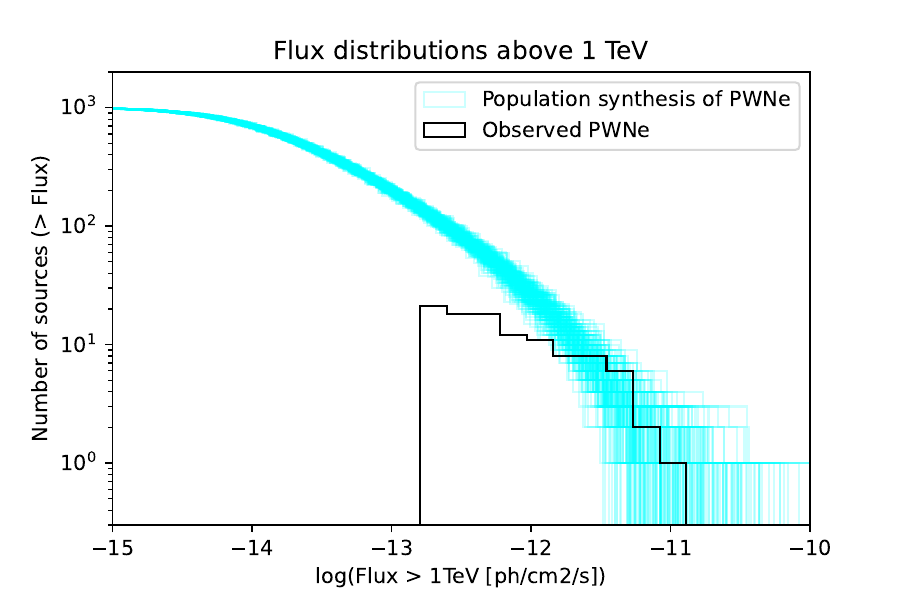}
    \caption{Flux distributions above 1 TeV of the full population of sources. The synthetic population (in blue) is directly compared with the firmly identified PWNe by \citet{2022A&A...666A...7M} (in black). The coloured areas represent the errors of the curve.}
    \label{fig:MW}
\end{figure}

In Figure~\ref{fig:MW}, we compare the expected  number distribution of TeV PWNe from the population synthesis to that of the observation. 
A summary of the parameters used to generate the population is reported in Table \ref{tabel:para_ngc253}, except that the index of lepton injection spectrum is form 1.75 to 2.25 randomly for PWN in Milky Way following \citet{2018A&A...612A...2H}. The electron conversion efficiency $\eta_e$ is fixed to 1 for all PWNe based on previous studies. Indeed, \citet{2018A&A...609A.110Z} modeled the dynamical and radiative evolution of 18 PWNe with a 1D leptonic model. Their results indicate  $0.93<\eta_e<0.99$ for six young PWNe ($T_{\rm age}<2400$~yr) and five evolved PWNe (2500~yr$<T_{\rm age}<$4600~yr), $\eta_e \sim 0.45$ for evolved PWNe MSH 15-52,
and $0.5<\eta_e<0.99$ for six mature/old PWNe ($T_{\rm age}>$6500~yr). Similar conclusions are found in other independent modeling of young PWNe, such as $\eta_e \in [0.6,0.99]$ by \citet{2014JHEAp...1...31T}, $\eta_e \in [0.75,0.95]$ by \citet{2011MNRAS.410..381B}, and $\eta_e$ approaching to 1 by \citet{2022ApJ...930L...2D}. 

We find that the $\log N - \log S$ distribution of the synthetic population is well consistent with that of the identified PWNe with TeV fluxes above $\sim 10^{-12}~\text{photons} /\text{cm}^2~\text{s}$. Below this value, the completeness of the observed sample drops due to limited sensitivity of instruments and hence the mocked PWN population outnumbers the observed one.  
The result is compatible to \citet{Fiori_2022}, \citet{2022A&A...666A...7M} and \citet{2020ApJ...904...85C}.
To put it shortly, our population model provides a satisfactory description of the currently observed PWN population with flux above $\sim 10^{-12}~\text{photons} /\text{cm}^2~\text{s}$. Among all the generated PWNe, young PWNe with age less than 10\,kyr contribute more than 90\% of the total TeV luminosity of the entire population. In other words, if all these PWNe are generated in an external galaxy (i.e., at roughly same distance from Earth), most the TeV flux emitted by the population would arise from young PWNe.

\section{TeV emission from PWN population in SBGs} \label{sec:apply}
In the following we will investigate whether a population of PWNe can reproduce the TeV emission in SBGs, and explore the parameter space of some key parameters of PWNe in SBGs. 

\subsection{Input parameters} \label{subsec:inSBN}

\begin{table*}
    \caption{Overview of parameter used in the modeling of PWNe.} 
    \label{tabel:para_ngc253}\setlength{\tabcolsep}{18pt}
    \begin{tabular}{lccc} 
      \hline 
      \textbf{Parameter} & \textbf{Symbol} & \textbf{Value}  & \textbf{\emph{varied} range}\\
      \hline
      \textbf{Particle spectrum at injection}\\
      Injection distribution index            & $\alpha$    & $\mathcal{U}(\alpha_\text{min},~\alpha_\text{min}+1)$ & \\
      Minimum index                           & $\alpha_\text{min}$  & & 1.5 ... 2.1        \\
      Injection distribution minimum energy   & $E_\text{min}$ (GeV)  & & 1 ... 1000  \\
      Injection distribution maximum energy    & $E_\text{max}$ (TeV)  & $\mathcal{U}(200,~800)$ &   \\
  
      \hline
      \textbf{PSRs population parameters} \\
      Braking index                 & $n$ & 3 & \\
      Initial spin periods          & $P_0$ (ms) & $\mathcal{N}(50,~35^2)(>10)$  &  \\
      Pulsar's equatorial surface Magnetic field  & log($B_{\rm s}$(G))   &  $\mathcal{N}(12.65,~0.55^2)$  &\\
      \hline
      \textbf{Nebulae} \\
      PWNe age limit                           & $\tau$(yr)      & $10^5$       &   \\
      Nebular magnetic field initial strength & $B_0$ ($\mu$G)   & $\mathcal{U}(50,~200)$  & \\
      Nebular magnetic field evolution index  & $\delta_B$      & 0.6 & \\
      \hline
        \multicolumn{4}{l}{ \textbf{NOTE ~--~}$\mathcal{U}(a,b)$ indicates a uniform distribution from \textit{a} to \textit{b}. } \\
        \multicolumn{4}{l}{~~~~~~~~~~~~~~~ $\mathcal{N}(\mu,\sigma^2)$ indicates a normal distribution with mean value $\mu$ and standard deviation $\sigma$. }
    \end{tabular}
\end{table*}

\begin{table}
    \caption{The energy density $U_{\text{Rad}}$ and the temperature $T$ of the three IR components due to dust and the optical one due to stars of NGC 253 and M82 from \citet{2019MNRAS.487..168P}. }
    \label{tabel:photonfield}
    \begin{tabular}{l|c|c} 
      \hline 
      \textbf{Photon field composition} & \textbf{NGC 253} & \textbf{M82}\\
      \hline
        $U_{\text{Rad}}^{\text{FIR}} \left(\text{eV}~ \text{cm}^{-3}\right)~ [T ~(\text{K})]$ & $1958~[40]$ & $910~[35]$ \\
        $U_{\text{Rad}}^{\text{MIR}} \left(\text{eV}~ \text{cm}^{-3}\right)~ [T ~(\text{K})]$ & $587~[101]$ & $637~[87]$\\
        $U_{\text{Rad}}^{\text{NIR}} \left(\text{eV}~ \text{cm}^{-3}\right)~ [T ~(\text{K})]$ & $587~[345]$ & $455~[278]$\\
        $U_{\text{Rad}}^{\text{OPT}} \left(\text{eV}~ \text{cm}^{-3}\right)~ [T ~(\text{K})]$ & $2936~[3858]$ & $546~[3829]$\\
      \hline
    \end{tabular}
\end{table}

We explore the influences of some main parameters of our model for each individual PWNe and the entire population.

\begin{description}
    \item  \verb'Supernova rate' --
    the number of pulsars is related to the rate of type-II SN explosions: $v_\text{SN}$. It determines the number of PWNe that are generated within $t_\text{end}$.  
    The estimation of $v_\text{SN}$ involves various methods, including stellar population fitting, line and FIR emission analysis, radio source modeling, and direct SN searches, but faces challenges due to star formation history, mass function variations, and uncertainties in direct SN searches \citep{2011ApJ...734..107L}. 
     $v_\text{SN}$ reported in the literature for M82 and NGC 253 span an order of magnitude, from $0.03~\text{yr}^{-1}$ to $0.3~\text{yr}^{-1}$ \citep{2018A&A...617A..73H, 2022ApJ...939..119B}.

    \item \verb|Conversion efficiency| -- 
    the fraction the spin-down energy going to electrons: $\eta_e$. The total spin-down power is usually considered to be divided into electrons injected into PWNe ($\eta_e$), magnetic fields ($\eta_B$) and other multi-messenger produced elsewhere from the PWNe ($\eta_\text{other}$) \citep{2009ApJ...703.2051G}. The following condition is naturally satisfied: $\eta_e+\eta_B+\eta_\text{other}=1$. Since $\eta_e$ and $v_\text{SN}$ are  linearly proportional to the amount of energy in relativistic electrons, these two parameters are degenerate.

    \item \verb|Energy spectrum of electrons| --
    the differential energy spectrum of the injected electrons can be described by a simple power-law function with index $\alpha$, ranging from $E_\text{min}$ to $E_\text{max}$.
    \citet{2022JHEAp..36..128M} summarized previous studies, and found that PWNe have very similar electrons injection index $\alpha \in [2.2,2.8]$ in Galaxy. We appropriately enlarge this range in the later simulation to better discuss the allowable parameter space.
    The energy range is relevant for the normalization of the spectrum and determines the ranges of the synchrotron and IC photon spectra.
    The maximum energy is related to the accelerations processes at the termination shock.
    Therefore, $E_{\rm max}$ for each PWN is randomly generated with a uniform distribution from 200\,TeV to 800\,TeV, following the treatment in \citet{2022A&A...666A...7M}. Its value would significantly influence the SED of the PWN for $\alpha \le 2$.
    On the contrary, the minimum $E_\text{min}$ will affect the resultant SED of the PWN for $\alpha > 2$. $E_{\rm min}$ is related to the Lorentz factor of the ultra-relativistic pulsar wind \citep{2000ApJ...539L..45C}. We consider $\alpha$ and $E_\text{min}$ as two free and independent parameters.

    \item \verb|Pulsar population| -- 
    a pulsar's initial rotational period $P_0$ and the surface magnetic field at the equatorial plane $B_{\rm s}$, as these determine the temporal evolution of particle injection. The initial period is particularly relevant because it sets the total available spin-down energy, $E_{\text{rot}}=2\pi^{2}I/P_{0}^{2}$ (with $I$ being the pulsar's moment of inertia).  
    We assume the distribution of parameters among new born pulsars in a SBG is similar to that in our Galaxy, and apply the PWN population generator that has been verified in the previous section to SBGs.
    
    \item \verb|Magnetic field| -- 
    the  strength of the initial magnetic field of PWNe, $B_0$, affects the synchrotron radiation and cooling of electrons. For a typical interstellar radiation of our Galaxy, young PWNe are expected to be synchrotron-dominated, with a low IC efficiency. However, the interstellar radiation environment in a SBN region is very different from that in our Galaxy.   As shown in the Table~\ref{tabel:photonfield}, the infrared radiation dominates the interstellar radiation of the SBN, and its energy density is more than three orders of magnitude higher than the typical value in our Galaxy, implying that the cooling time of highly-relativistic electrons is much shorter. For an electron with energy of 1 TeV, the cooling timescale is approximately 200~yrs in SBN, while it is $3 \times 10^5~\text{yrs}$ in ISM of the Milky Way. The IC cooling in such an environment will dominate over synchrotron losses unless $B \gtrsim 500~\mu \text{G}$.
    For typical values of  $B_0\approx 100~\mu \text{G}$, as found in some young PWNe \citep[e.g.][]{2012arXiv1202.1455M, 2018A&A...612A...2H}, we expect a large fraction of the injected electron energy to be channeled into the IC radiation at GeV - TeV energies.

\end{description}

In Table~\ref{tabel:para_ngc253}, 
we summarize the parameters and relative distribution (or the allowed range) used as initial condition for the simulation of the PWNe population.

The target photon fields in SBG include CMB, infrared, and optical radiation fields, where the latter two in SBG are much stronger than those in our Galaxy.
\citet{2019MNRAS.487..168P} obtain energy densities and  temperatures of the three IR components arising from dust emission and those of the optical radiation from stars. 
The parameters of these radiation fields for NGC~253 and M82 are listed in Table \ref{tabel:photonfield}, which are employed in our following calculations.

\subsection{NGC~253} \label{subsec:ncg253}

NGC~253 is one of the only two starburst galaxies found to emit $\gamma$-ray from hundreds of MeV \citep{2010ApJ...709L.152A} to multi-TeV energies \citep{2018A&A...617A..73H} . 
Based on the planetary nebula luminosity function, a weighted average of the most reliable distance estimates yields a distance of $d=3.5~\pm~0.2~\text{Mpc}$ by \citet{2005MNRAS.361..330R}.

\citet{2002ApJ...574..709M} derive a star-formation rate of $\sim 3.5 M_{\sun} \text{yr}^{-1}$, based on the far-infrared luminosity in the starburst nucleus of NGC 253. \cite{2013MNRAS.429L..70O} further estimate a type-II SN rate $v_\text{SN}$ in the starburst nucleus of NGC 253 to be 0.02~$\text{yr}^{-1}$, whereas \cite{2018A&A...617A..73H} suggest an SN rate within the starburst region of NGC 253 of $v_\text{SN} \approx 0.05~\text{yr}^{-1}$. We take it as a reference pulsar birth rate, and generate 5000 pulsars in our sample since we consider emission of PWNe with ages up to 100\,kyr. 


\begin{figure}
    \includegraphics[width=\columnwidth]{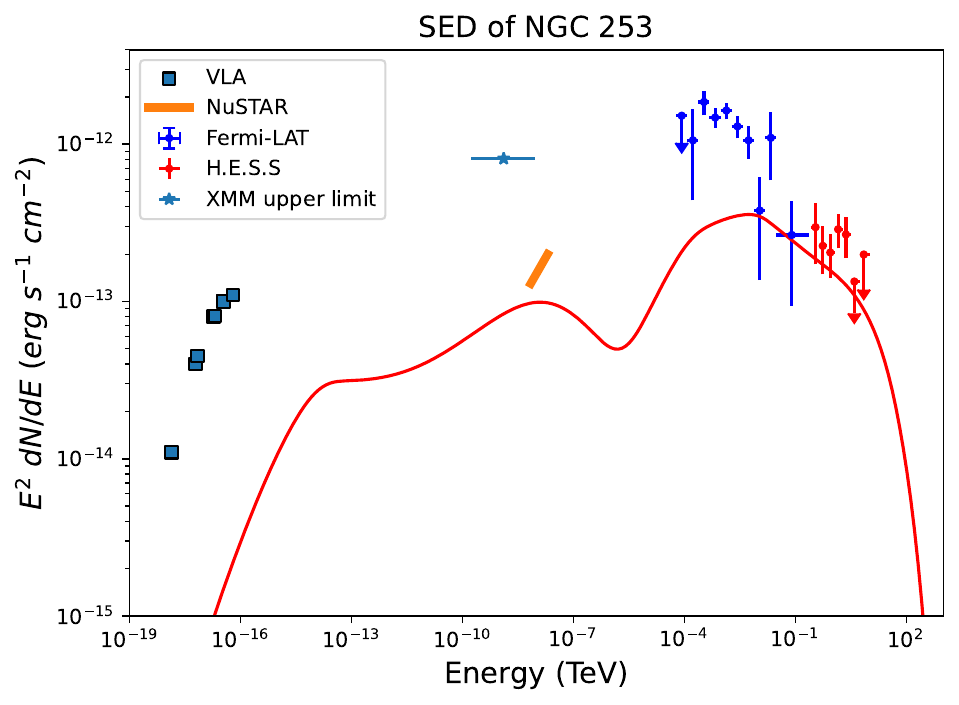}
    \caption{Predicted  SED for PWNe populations in the starburst region of NGC 253 according to baseline model. 
    Also shown is radio data from the Very Large Array \citep{1996A&A...305..402C}, the XMM upper limit is emission from the central source X 34 as given by \citet{2001A&A...365L.174P}, the hard X-ray upper limit is taken from \citet{2014ApJ...797...79W}, the \textit{Fermi}-LAT and H.E.S.S. data is from \citet{2012ApJ...757..158A}. Horizontal error bars show the energy band over which a particular observation is made, while vertical error bars show $1\sigma$ uncertainties; the upper limits are given at 95\% confidence level. }
\label{fig:SED_ngc}
\end{figure}

For each of the simulated pulsars, we assign the values of $P_0$ and $B_{\rm s}$ to them following the description in Section~\ref{sec:MW}. We assume the same values of $\eta_e$ and $E_{\rm min}$ for each of the corresponding PWN, and randomly set the value of their injection spectral index $\alpha$ in the range between $\alpha_{\rm min}$ and $\alpha_{\rm min}+1$ with an equal probability. 
To explore the effect of key parameters $\alpha$, $E_\text{min}$ and $\eta_e \times v_\text{SN}$ on emission of their PWNe, we test 140 ($=7\times20$) different combinations of values of $\alpha_{\rm min}$ and $E_{\rm min}$, where $\alpha_\text{min}=[1.5,~2.1]$ with a linear increment of 0.1 and $E_\text{min}=[1,~1000]~\text{GeV}$ with a logarithmic increment of 0.05 dex. The value of $\eta_e \times v_\text{SN}$ affect the fluxes of PWNe linearly and can be determined by matching the observed flux of the galaxy.

In Figure~\ref{fig:SED_ngc}, the red line shows the SED produced by a population of PWNe in NGC 253 with the so-called "baseline" model parameters, i.e. $ \alpha \in \mathcal{U}(1.8,~2.8)$ and $E_\text{min}=162$ GeV.  
Based on the analysis in Section~\ref{subsec:inSBN}, we choose $\eta_e \times v_\text{SN} = 0.05~\text{yr}^{-1}$ as an intermediate parameter to analyze.
In this baseline  model, the IC emission can explain the TeV emission of NGC 253 while the synchrotron emission does not exceed the X-ray limit. In the figure, the \textit{NuSTAR} upper limit is obtained by subtracting the two known hard X-ray components from the observed flux:  thermal gas and X-ray binaries point sources \citep{2014ApJ...797...79W} .
The remaining represents the unresolved, diffuse non-thermal emission from the galaxy.

\begin{figure}
    \includegraphics[width=\columnwidth]{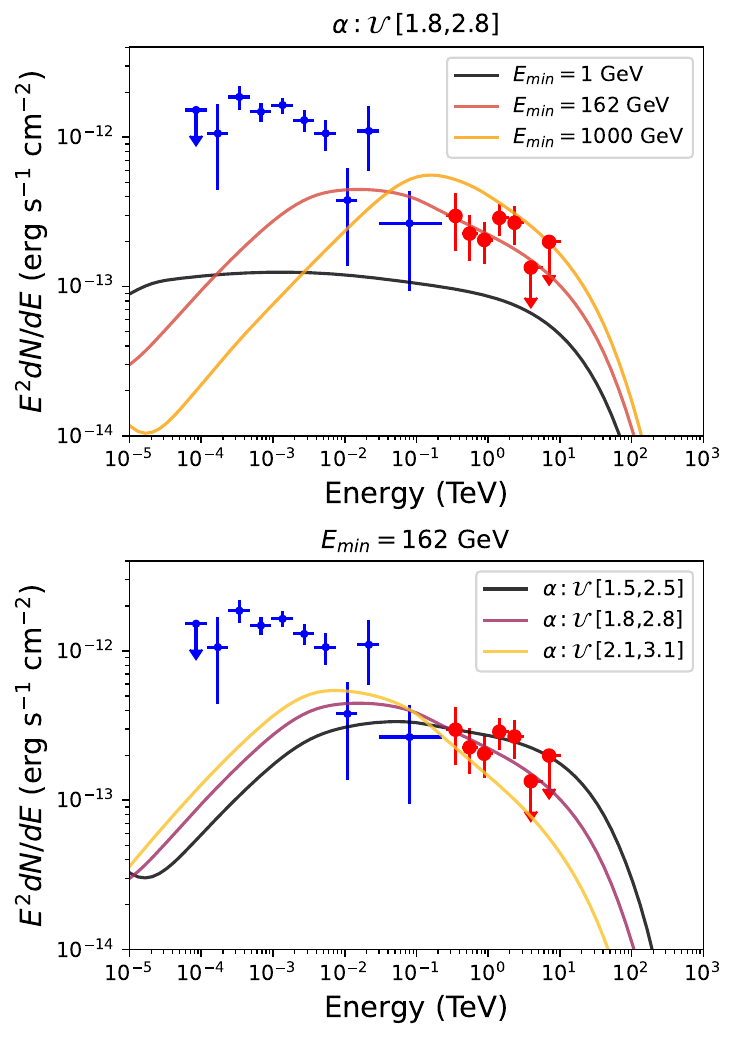}
    \caption{Predicted SED for PWNe populations in the starburst region of NGC 253. $Fermi$-LAT points are dots (blue), and H.E.S.S. points are circle (red). Single case flux calculated by changing the value of the two most relevant parameters, one for each panel. }
\label{fig:SED_ngc2}
\end{figure}

\begin{figure*}
    \includegraphics[width=2\columnwidth]{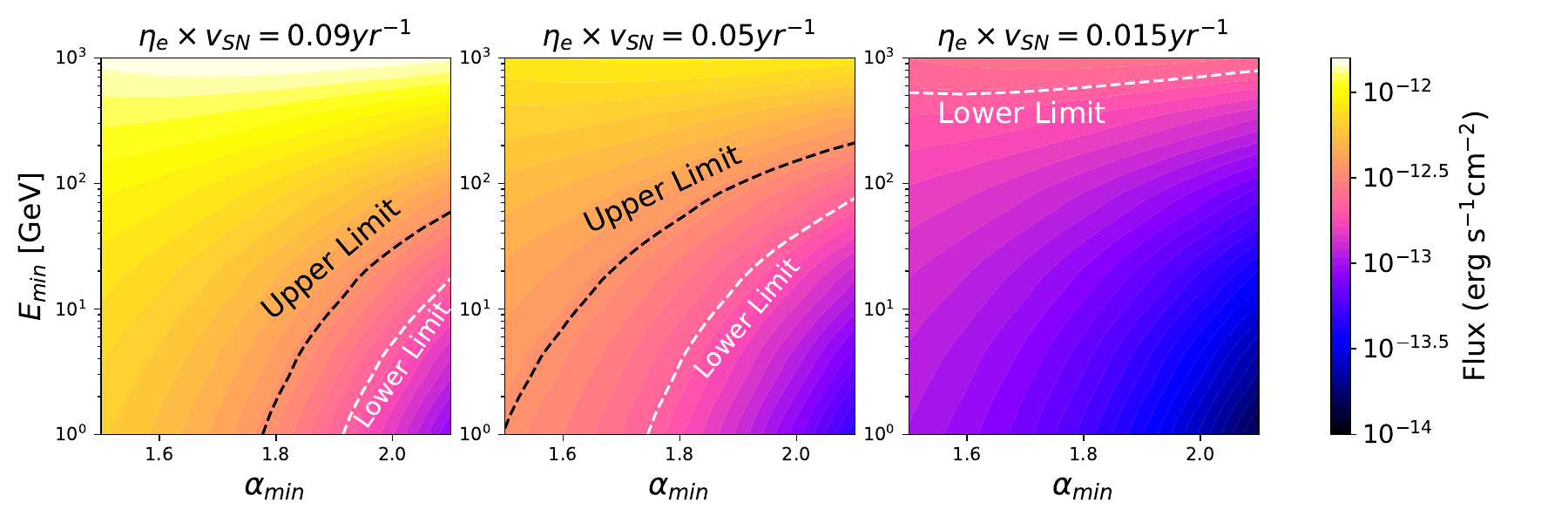}
    \caption{For NGC 253, allowable parameter space for $\alpha_\text{min}$, $E_\text{min}$ and $\eta_e \times v_\text{SN}$. Deriving parameter ranges by limiting flux at 1 TeV with $1\sigma$ statistical uncertainty by H.E.S.S. The blue and white dashed lines represent the upper and lower limits of the 1$\sigma$ respectively, and the middle area is the allowable parameter space.}
    \label{fig:eta}
\end{figure*}

We investigate the dependence of the gamma-ray flux on different values of $\alpha_\text{min}$ and $E_\text{min}$. While varying one of these two parameter, we fix other parameters the same as those in the baseline model.  By summing up the emission of each PWN, we obtain the SED of NGC~253, as shown in the two panels of Figure~\ref{fig:SED_ngc2}. 
The top panel of Fig.~\ref{fig:SED_ngc2} illustrates the impact of changing $E_\text{min}$, from 1 GeV to 1000 GeV. The figure shows that the gamma flux is sensitive to the minimum energy truncation of accelerated electrons: the larger minimum energy truncation, the higher flux.
The bottom panel of Fig.~\ref{fig:SED_ngc2} illustrates the dependence of the results on the injection spectral index, from $\mathcal{U}(1.5,~2.5)$ to $\mathcal{U}(2.1,~3.1)$.
As is shown, a steeper spectrum leads to a softer gamma-ray spectrum and a smaller gamma-ray flux above 0.1\,TeV.


The parameter space able to account for the TeV flux of NGC 253 is shown in Figure \ref{fig:eta}, which is obtained by
comparing the model and the observed flux at 1~TeV  within $1\sigma$ statistical uncertainties \citep{2018A&A...617A..73H}. Combination of $E_{\rm min}$ and $\alpha_{\rm min}$ in the upper left side of the dashed black curve results in a too high TeV flux while that in the lower right side the dashed white curve results in a too low TeV flux. Therefore, the region between the black and white dashed curves are available parameter space. Three panels are obtained with different values of $\eta_e\times v_{\rm SN}$. We see that the available region is broad and covers the typical values that are usually considered for PWNe, as long as the value of $\eta_e\times v_{\rm SN}$ is not too small. When $\eta_e \times v_\text{SN} \lesssim 0.015 {\rm yr}^{-1}$, 
there will be almost no parameter space that can match the observed TeV flux. Given $v_{\rm SN}=0.05\,\rm yr^{-1}$ as the typical type-II SN rate in the starburst galaxy, it requires PWNe to have a relatively high pair conversion efficiency of $\eta_e>0.3$ to make a significant contribution to the TeV radiation of the galaxy.
Such a high $\eta_e$ is actually common for young PWNe ($T_{\rm age}<$ 10~kyr) in Milky Way according to previous studies, as discussed in Section \ref{sbusec:gamma}.
Results of these studies support the PWN-origin of the TeV emission from starburst galaxies.
 
\subsection{M82}  \label{subsec:m82}

M82, at a distance of $d=3.6~\pm~0.3~\text{Mpc}$, is also detected in GeV and TeV $\gamma$-ray by \textit{Fermi}-LAT  and VERITAS, respectively \citet{2010ApJ...709L.152A,2009Natur.462..770V}.  
The inferred SFR is $\sim 10 M_{\sun} \text{yr}^{-1}$, considerably higher than that in NGC 253 and \citet{2013MNRAS.429L..70O} estimate a type-II SN rate of 0.06~$\text{yr}^{-1}$. 
\cite{2010ApJ...709L.152A} estimated that the SN explosion rate varies from $\approx 0.08$ to $0.3~\text{yr}^{-1}$ in M82. Similar to NGC 253, we also consider $\eta_e \times v_\text{SN}$ of M82 as a combined free parameter from 0.01 to 0.1 yr$^{-1}$.

\begin{figure}
    \includegraphics[width=\columnwidth]{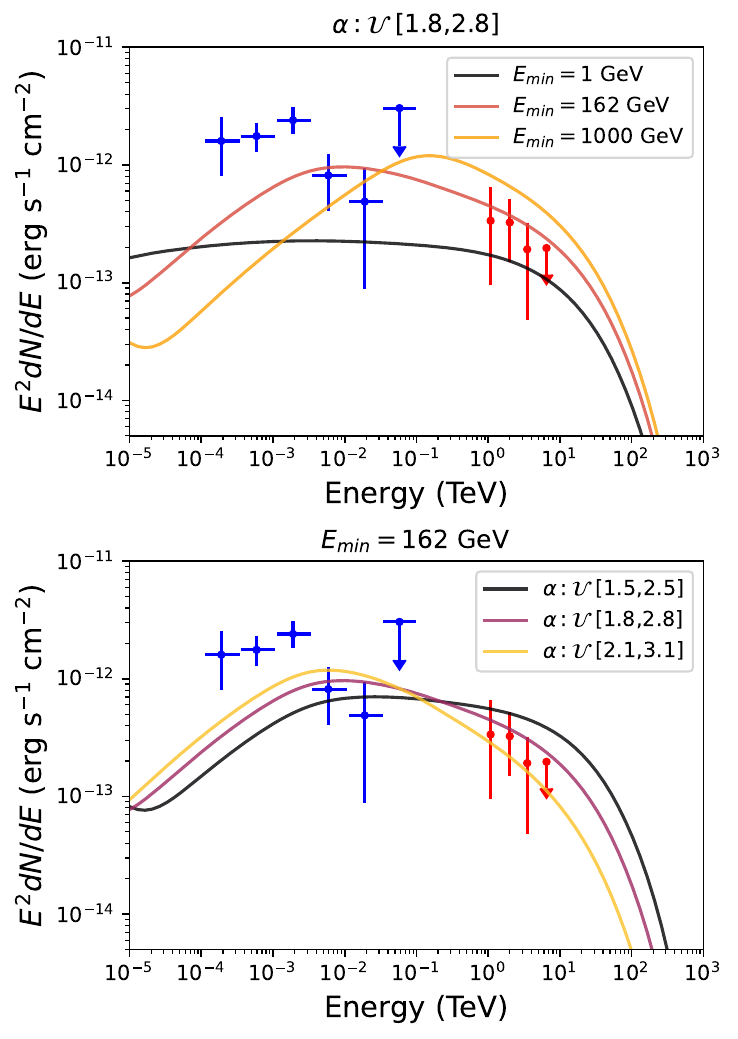}
    \caption{Same as Figure \ref{fig:SED_ngc2}, but for the SBG M82. 
     Fermi-LAT points are blue crosses \citep{2012ApJ...755..164A}, and VERITAS points are red circles \citep{2009Natur.462..770V}. }
    \label{fig:M82}
\end{figure}

The calculated SED has very similar properties to those of  NGC 253, as expected due to the similar target radiation field, magnetic fields, and average particle densities in the SB regions (see Figure~\ref{fig:M82}). 

\begin{figure*}
    \includegraphics[width=2\columnwidth]{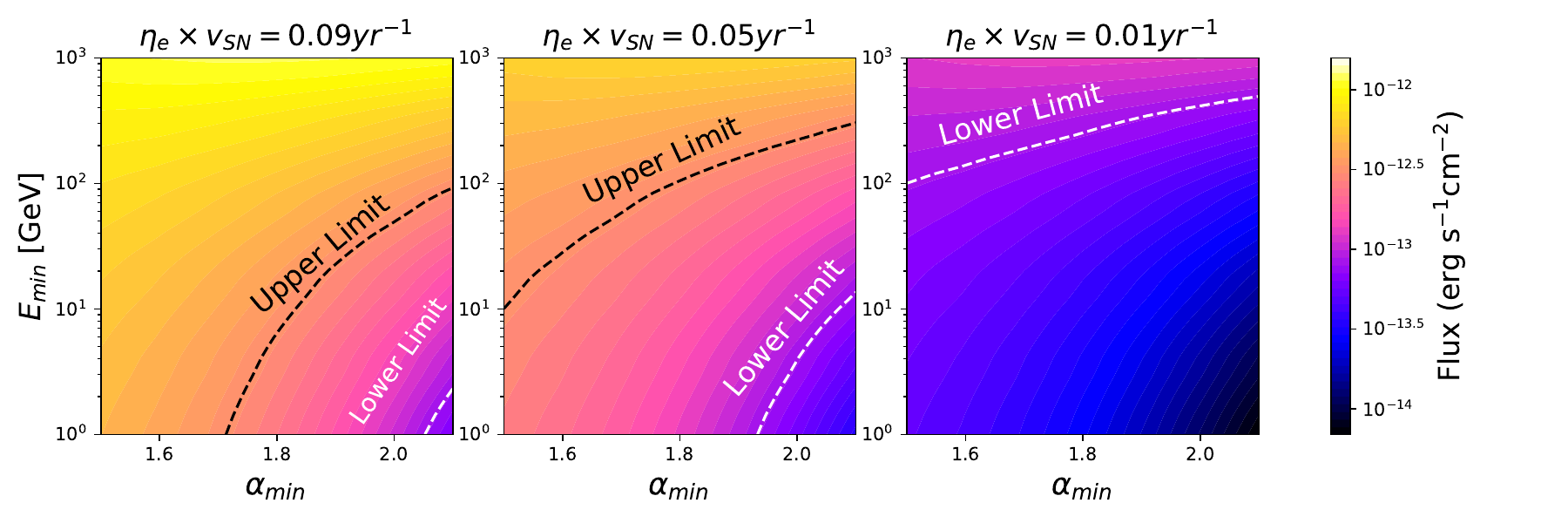}
    \caption{Allowable parameter space for M82, as Figure \ref{fig:eta}. 
    The blue and white dashed lines represent the upper and lower limits of the 1$\sigma$ by \textit{VERITAS} respectively, and the middle area is the allowable parameter space. }
\label{fig:eta_M82}
\end{figure*}

The parameter space of PWNe to account for the TeV flux of  M82 is given in Figure~\ref{fig:eta_M82}. The available region in the parameter space for M82 is broader compared to that for NGC 253, because VERITAS observation gives a larger statistical error for the TeV flux. When $\eta_e \times v_\text{SN} \la 0.01~{\rm yr}^{-1}$, we find that there will be very limited parameter space that can match the observed TeV radiation, as can be see from Figure~\ref{fig:eta_M82}.

\subsection{Constraint on initial magnetic field strength of PWNe}  \label{subsec:B_PWN}

Although the initial magnetic field strength of PWNe dose not affect TeV emission of SBG (see Section~\ref{subsec:inSBN}), it can affect the intensity of synchrotron radiation, so the X-ray observations can be used to constrain the initial magnetic field of PWNe.

\citet{2014ApJ...797...79W} utilize the \textit{NuSTAR} and \textit{Chandra} data to investigate the populations contributing to the galaxy-wide 0.5-30 keV emission from NGC 253. Subtracting  the contribution of resolved  sources and contribution from diffuse gas thermal emission, they determine the 90\% upper limit on the non-thermal flux in the 7-20 keV band. Hard X-ray are mainly produced by synchrotron radiation of TeV electrons, according to $\nu_c = 3eB\gamma_e^2/4\pi m_ec$. This limit constrains that the initial magnetic field of PWNe cannot be too high, otherwise the synchrotron  emission will exceed the upper limit.

\begin{figure*}
    \includegraphics[width=2\columnwidth]{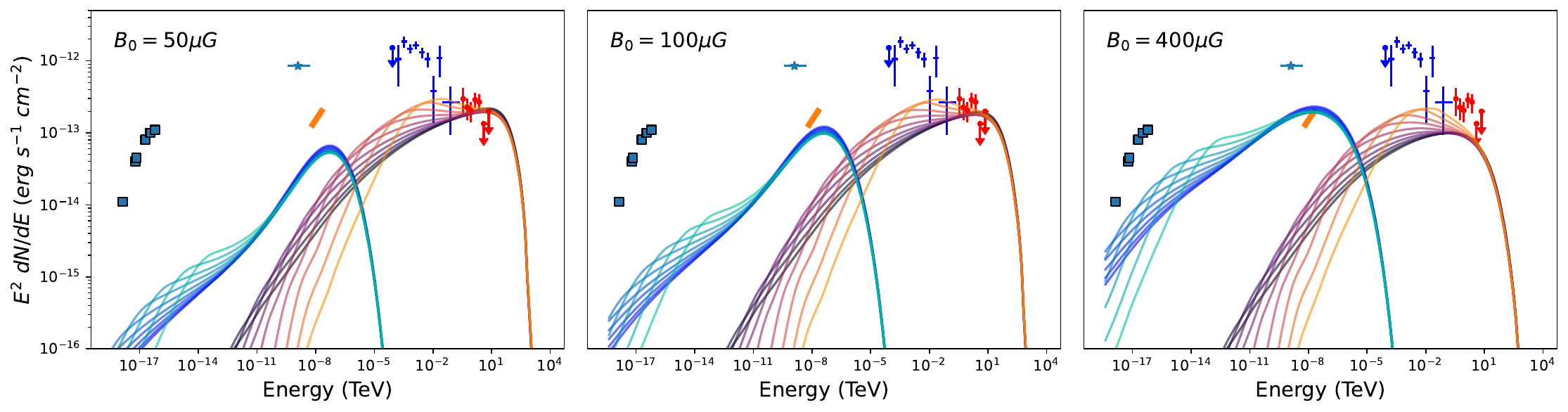}
    \caption{Same as Figure~\ref{fig:SED_ngc}, \textit{Fermi}-LAT points are dots (blue), H.E.S.S. points are circle (red), \textit{NuSTAR} upper limit is line (orange), \textit{Chandra} total radiation is star (cyan), Very Large Array points are square. The blue-purple lines with lower energy represent synchrotron radiation, and the red-black lines with higher energy represent IC radiation. In each panel, taking $ \alpha \in \mathcal{U}(1.6,~2.6)$ as an example, the color lines from dark to light indicate the $E_\text{min}$ cases from from 1MeV to 100\,GeV on the logarithmic interval, respectively.}
\label{fig:Xray}
\end{figure*}

Fig.~\ref{fig:Xray} shows the SED under different initial magnetic fields of PWNe. 
Due to that the energy spectrum is sensitive to the electron injection spectrum,  we take $ \alpha \in \mathcal{U}(1.6,~2.6)$ as an example. In each panel, the color lines from dark to light indicate the $E_\text{min}$ increases from  1 MeV to 100\,GeV on the logarithmic interval, respectively. This also proves that the minimum energy truncation $E_\text{min}$ does not affect the hard X-ray flux.

The initial nebular magnetic field strength in PWNe effects the part of non-thermal electrons energy channeled into high-energy photon for by synchrotron radiation.
Based on the allowable parameter space (Fig.~\ref{fig:eta}) given by the limitation of the TeV band, when the magnetic field exceeds 400~$\mu$G quantitatively,  the flux in the hard X-ray band will exceed the upper limit given by the X-ray observations, so we obtain an upper limit of $B_0 < 400 \mu$G for the initial magnetic field of PWNe in starburst galaxy.

\section{Discussions} \label{discussion}

We here discussion the influence of some subordinate parameters on the results. 

\begin{description}
    \item  \textbf{The effect of the maximum age for PWNe} -- The pulsar age is relevant for the particle accumulation, dynamics and energetic. We have verified, however, that varying the final age of PWNe does not affect the results of the present study, given that the $\gamma$-ray luminosity of the associated PWNe powered by older pulsars is negligible. The number of pulsars with age $t_{\rm age}$ is proportional to $t_{\rm age}$ given a constant pulsar birth rate. On the other hand, the spin-down luminosity decreases with $t_{\rm age}^{-2}$ for a braking index of 3. Therefore, the total spin-down luminosity of pulsars of age $t_{\rm age}$ at the present time roughly scales with $t_{\rm age}^{-1}$.  
    Note that due to the intense infrared radiation field in starburst galaxies, the cooling timescale of the high energy electron is quite short. Therefore, electrons injected at early stage, when the spin-down luminosity of those middle-aged pulsars were high, cannot survive at the present time. In addition, for middle-aged and old pulsars, the average electron conversion efficiency $\eta_e$ is probably only at the level of 0.1, as constrained by the diffusive $\gamma$-ray emission from the Galactic plane \citep{2023PhRvD.107j3028Y}. Therefore, the contribution of older pulsars is negligible.
    
    
   \item \textbf{Influence of the magnetic field of ISM} -- The pulsar may escape the parent SNR shell at a time $t_\text{cross} \simeq 45 (E_\text{SN} /  10^{51}~\text{erg})^{1/3}(n_\text{ISM} / 1~\text{cm}^{-3})^{-1/3} (v_p/400~\text{km~s}^{-1})^{-5/3} \text{kyr}$ \citep{2011piim.book.....D} due to its high kick velocity. The high-energy electrons can largely escape outer of the PWNe and diffuse into the surrounding ISM, producing the so called “pulsar halos”, as found recently in the Milky Way  \citep{HAWC_halo,LHAASO_0622}.    
   Unlike the Milky Way,  SBN is characterised by a much higher magnetic field and average density ($B_\text{ISM} \sim 250~\mu \text{G}, \ n_\text{ISM} \sim 250~\text{cm}^{-3}$), so the synchrotron losses may be stronger than the IC loss for these high-energy electrons that have escaped the PWNe. Therefore, the contribution to the TeV emission of SBGs by pulsar halos may be subdominant. 

   \item \textbf{Influence of the magnetic field at the beginning stage} -- As $t<\tau_0(\sim 0.5 \rm kyr)$, the employed evolution of the magnetic field, i.e., Eq.~\ref{equation_B} suggests a constant magnetic field at the beginning. At this stage, the PWN expands freely and its radius increase with time as $R_\text{PWN} \propto t^{6/5}$ \citep{2006ARA&A..44...17G}.
   If assuming magnetic flux conservation in the PWN \citep{2018A&A...612A...2H}, the magnetic field strength would evolve as $B(t) \propto t^{-12/5}$. For a given magnetic field as revealed from the SED of a PWN at the present time, it would imply a much stronger magnetic field strength at the beginning. However, even for the early magnetic field predicted by Eq.~\ref{equation_B}, relativistic electrons responsible for TeV emission cool very rapidly via the synchrotron radiation. Therefore, those electrons injected at early time cannot survive at the present time in either case. 
   For example, in the PWN of SN 1986J ($t \sim 30~\text{yr},\ L\sim 10^{39} ~\text{erg~s}^{-1},\ B_\text{PWN} \sim 17~\text{mG}$, 
   \citealt{2023MNRAS.525.2750T}) , the cooling timescale of relativistic electrons is about $20(E/1\rm TeV)^{-1}\,$days given the inferred magnetic field. Considering a two-segmented magnetic field evolution scenario, it would affect the resulting gamma-ray spectra at most in the order of $\sim 10\%$,  which is negligible compared to parameters discussed in the previous section.

    { \item \textbf{Effect of braking index} -- 
    While our calculations are based on the assumption of braking index $n=3$ for all generated pulsars, some of detected pulsars present different braking indexes, such as 2.5 for the Crab pulsar. The braking index influences the spin-down history. However, taking a different braking index would not affect our result significantly, mainly due to two reasons. First, the property of the simulated pulsar population need to match that of the observed pulsar sample, for example, in terms of distributions of $\dot{P}$ and $P$. As a result, the distribution of the spin-down luminosity of pulsars, which is determined by $-\dot{P}/P^3$, is more or less the same, regardless of the chosen value of the braking index. Consequently, the total electron injection luminosity in their PWNe is insensitive to the braking index. On the other hand, the gamma-ray luminosity at the present day depends on all the cumulative  electrons injected in the history, which need be traced back over a time period equal to the cooling timescale of emitting electrons $t_c$. Given that pulsar's spin-down luminosity evolves with time $t$ as
    \begin{equation}
        L(t) = L_0 \left( 1+\frac{t}{\tau_0}\right)^{-\frac{n+1}{n-1}},
    \end{equation}
    the difference, i.e., the ratio, between the electron injection luminosity at the present day ($t=t_{\rm age}$) and a period of time $t_c$ ago  ($t=t_{\rm age}-t_c$) can be given by $f=\left( \frac{1+t_{\rm age}/\tau_0}{1+(t_{\rm age}-t_c)/\tau_0}\right)^{-(n+1)/(n-1)}$. Given that the magnetic field inside the PWN also evolve with time, we estimate the cooling timescale by only considering the IC cooling for simplicity. In the environment of the starburst nucleus, the typical IC cooling timescale for TeV-emitting electron is about $t_c=200\,$yr. $\tau_0$ of most simulated pulsars range in $800-5000\,$yr. We find that the ratio  $f$ is around $\sim 0.5 - 1 $. As such, changing $n=3$ to $n=2$ only alter the value of $f$ by a factor less than 2. Indeed, due to the rapid cooling of TeV-emitting electrons in the environment of starburst nucleus, electrons can be only accumulated over a short period of time. As a result, the total amount of emitting electrons does not rely on the braking index which controls the injection history. Instead, it is basically determined by the present-day spin-down luminosity of each pulsar, the distribution of which is calibrated by the observed pulsar sample. 
    In Fig.\ref{fig:braking}, we show the result with different braking index and we find that the difference in the resulting flux is less than 20\%.}
   
\end{description}

\begin{figure}
    \includegraphics[width=\columnwidth]{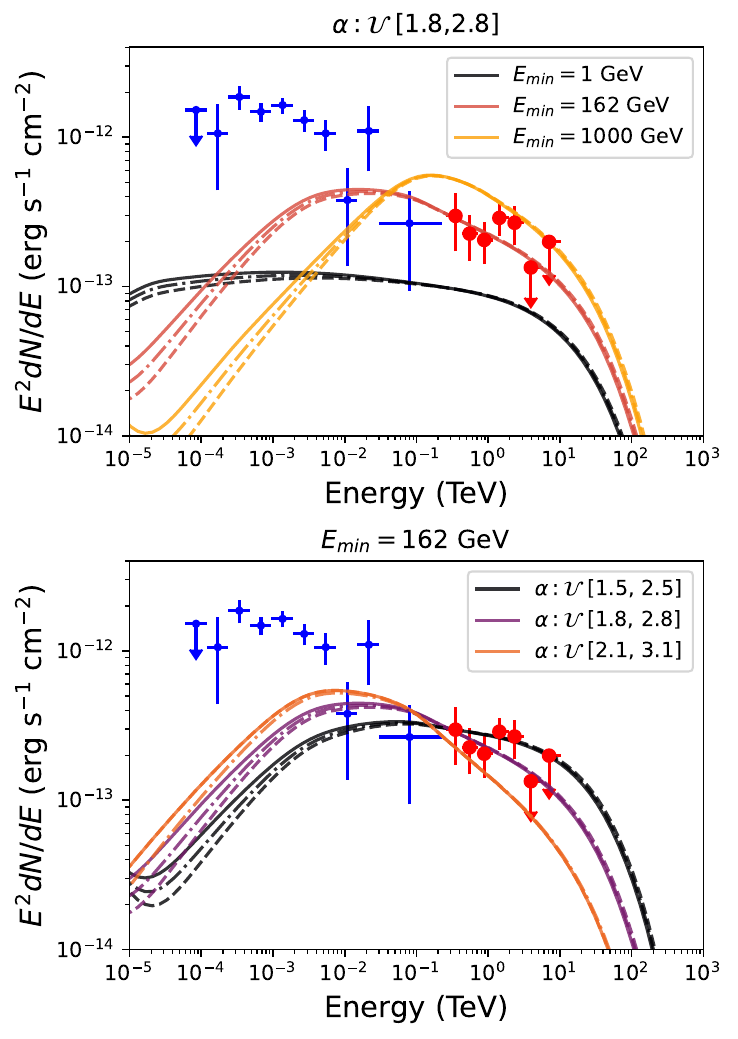}
    \caption{The impact of different braking index on the final results. The solid lines represent braking index n = 3, the dot–dashed lines represent n = 2.5, and the dashed lines represent n = 2. }
    \label{fig:braking}
\end{figure}

\section{Conclusion} \label{conclusions}

Starburst galaxies such as M82 and NGC 253 show a harder high-energy $\gamma$-ray spectrum with a higher luminosity than that of Milky Way. While it is generally considered that the TeV $\gamma$-ray emission of the starburst galaxies arise from interactions of injected cosmic ray hadrons with interstellar medium, we found that PWNe in starburst regions may also explain the observed TeV flux. 

To model the PWN population and subsequently their emission in the starburst galaxies, we firstly simulate the PWN population in Milky Way for verification of our method. We followed the statistical study of properties of pulsars and PWNe in the previous literature and found that the simulated $\log N-\log S$ distribution of TeV flux of PWNe in Milky Way is consistent with observations. 
We then applied the method to the starburst galaxies NGC~253 and M82, and calculate the spectral energy distribution produced by the PWN population in the SBGs. 
Our main results are as follows:
\begin{enumerate}
    \item PWNe associated with core-collapse supernovae in starburst regions may explain the observed TeV emission of SBG NGC~253 and M82 with typical parameters which are usually employed in previous literature for PWNe. 

    \item From the perspective of energy budget, we found that it generally requires $\eta_e\times v_{\rm SN}>0.01\, \rm yr^{-1}$ for the PWN population and the galaxy. 
    
    \item By requiring the synchrotron emission of the PWNe not to exceed the hard X-ray observations of SBGs, we constrain the initial magnetic field strength of PWNe to be less than $400\,\mu$G.
    
\end{enumerate}

If GeV and TeV emission of SBGs come from the hadronic radiation of CRs in ISM and the leptonic emission of PWNe respectively, we would expect a spectral break somewhere between GeV and TeV energy. 
Future observations on SBGs by sensitive gamma-ray instruments such as LHAASO \citep{2019arXiv190502773C} and the Cherenkov Telescope Array (CTA) \citep{2021MNRAS.506.6212S} is potential to measure such a feature in the spectra of SBGs, which would serve as a critical test of the scenario. Besides, the next-generation neutrino instruments, such as IceCube-Gen2 \citep{2021JPhG...48f0501A}, could detect neutrinos from SBGs with a long-term exposure if the TeV emission is dominated by hadronic process \citep{2021ApJ...907...26H}. Thus in combination with future neutrino and gamma-ray observations, the origin of leptonic or hadronic emission from SBGs may be distinguished.

\section*{Acknowledgements}
This study is supported by National Scientific Foundation of China, under grants No. U2031105, No.12121003, No. 12333006, and China Manned Spaced Project (CMS-CSST-2021-B11).

\section*{Data Availability}
The data of pulsars are available at the ATNF pulsar catalog, at {\url{https://www.atnf.csiro.au/people/pulsar/psrcat/}}. 
For the M82, the Fermi-LAT data is taken from \citet{2012ApJ...755..164A}, at {\url{https://dx.doi.org/10.1088/0004-637X/755/2/164}}, 
and the VERITAS data is taken from \citet{2009Natur.462..770V}, at {\url{https://doi.org/10.1038/nature08557}}.
For the NGC 253, the VLA data is taken form \citep{1996A&A...305..402C}, at {\url{https://ui.adsabs.harvard.edu/abs/1996A%26A...305..402C}},
the NuSTAR data is taken from \citet{2014ApJ...797...79W}, at {\url{http://dx.doi.org/10.1088/0004-637X/797/2/79}}, 
the XMM data is taken from \citet{2001A&A...365L.174P}, at {\url{http://dx.doi.org/10.1051/0004-6361:20000068}}, 
and the H.E.S.S data and Fermi-LAT data is taken from \citet{2012ApJ...757..158A}, at {\url{http://dx.doi.org/10.1088/0004-637X/757/2/158}}.

\bibliographystyle{mnras}
\bibliography{mybib} 

\begin{thebibliography}{}
\makeatletter
\relax
\def\mn@urlcharsother{\let\do\@makeother \do\$\do\&\do\#\do\^\do\_\do\%\do\~}
\def\mn@doi{\begingroup\mn@urlcharsother \@ifnextchar [ {\mn@doi@} {\mn@doi@[]}}
\def\mn@doi@[#1]#2{\def\@tempa{#1}\ifx\@tempa\@empty \href {http://dx.doi.org/#2} {doi:#2}\else \href {http://dx.doi.org/#2} {#1}\fi \endgroup}
\def\mn@eprint#1#2{\mn@eprint@#1:#2::\@nil}
\def\mn@eprint@arXiv#1{\href {http://arxiv.org/abs/#1} {{\tt arXiv:#1}}}
\def\mn@eprint@dblp#1{\href {http://dblp.uni-trier.de/rec/bibtex/#1.xml} {dblp:#1}}
\def\mn@eprint@#1:#2:#3:#4\@nil{\def\@tempa {#1}\def\@tempb {#2}\def\@tempc {#3}\ifx \@tempc \@empty \let \@tempc \@tempb \let \@tempb \@tempa \fi \ifx \@tempb \@empty \def\@tempb {arXiv}\fi \@ifundefined {mn@eprint@\@tempb}{\@tempb:\@tempc}{\expandafter \expandafter \csname mn@eprint@\@tempb\endcsname \expandafter{\@tempc}}}

\bibitem[\protect\citeauthoryear{{Aartsen} et~al.,}{{Aartsen} et~al.}{2021}]{2021JPhG...48f0501A}
{Aartsen} M.~G.,  et~al., 2021, \mn@doi [Journal of Physics G Nuclear Physics] {10.1088/1361-6471/abbd48}, \href {https://ui.adsabs.harvard.edu/abs/2021JPhG...48f0501A} {48, 060501}

\bibitem[\protect\citeauthoryear{{Abdo} et~al.,}{{Abdo} et~al.}{2010}]{2010ApJ...709L.152A}
{Abdo} A.~A.,  et~al., 2010, \mn@doi [\apjl] {10.1088/2041-8205/709/2/L152}, \href {https://ui.adsabs.harvard.edu/abs/2010ApJ...709L.152A} {709, L152}

\bibitem[\protect\citeauthoryear{{Abeysekara} et~al.,}{{Abeysekara} et~al.}{2017}]{HAWC_halo}
{Abeysekara} A.~U.,  et~al., 2017, \mn@doi [Science] {10.1126/science.aan4880}, \href {https://ui.adsabs.harvard.edu/abs/2017Sci...358..911A} {358, 911}

\bibitem[\protect\citeauthoryear{{Abramowski} et~al.,}{{Abramowski} et~al.}{2012}]{2012ApJ...757..158A}
{Abramowski} A.,  et~al., 2012, \mn@doi [\apj] {10.1088/0004-637X/757/2/158}, \href {https://ui.adsabs.harvard.edu/abs/2012ApJ...757..158A} {757, 158}

\bibitem[\protect\citeauthoryear{{Ackermann} et~al.,}{{Ackermann} et~al.}{2012}]{2012ApJ...755..164A}
{Ackermann} M.,  et~al., 2012, \mn@doi [\apj] {10.1088/0004-637X/755/2/164}, \href {https://ui.adsabs.harvard.edu/abs/2012ApJ...755..164A} {755, 164}

\bibitem[\protect\citeauthoryear{{Aharonian} et~al.,}{{Aharonian} et~al.}{2021}]{LHAASO_0622}
{Aharonian} F.,  et~al., 2021, \mn@doi [\prl] {10.1103/PhysRevLett.126.241103}, \href {https://ui.adsabs.harvard.edu/abs/2021PhRvL.126x1103A} {126, 241103}

\bibitem[\protect\citeauthoryear{{Arzoumanian}, {Chernoff}  \& {Cordes}}{{Arzoumanian} et~al.}{2002}]{2002ApJ...568..289A}
{Arzoumanian} Z.,  {Chernoff} D.~F.,   {Cordes} J.~M.,  2002, \mn@doi [\apj] {10.1086/338805}, \href {https://ui.adsabs.harvard.edu/abs/2002ApJ...568..289A} {568, 289}

\bibitem[\protect\citeauthoryear{{Behrens} et~al.,}{{Behrens} et~al.}{2022}]{2022ApJ...939..119B}
{Behrens} E.,  et~al., 2022, \mn@doi [\apj] {10.3847/1538-4357/ac91ce}, \href {https://ui.adsabs.harvard.edu/abs/2022ApJ...939..119B} {939, 119}

\bibitem[\protect\citeauthoryear{{Blumenthal} \& {Gould}}{{Blumenthal} \& {Gould}}{1970}]{1970RvMP...42..237B}
{Blumenthal} G.~R.,  {Gould} R.~J.,  1970, \mn@doi [Reviews of Modern Physics] {10.1103/RevModPhys.42.237}, \href {https://ui.adsabs.harvard.edu/abs/1970RvMP...42..237B} {42, 237}

\bibitem[\protect\citeauthoryear{{Bucciantini}, {Arons}  \& {Amato}}{{Bucciantini} et~al.}{2011}]{2011MNRAS.410..381B}
{Bucciantini} N.,  {Arons} J.,   {Amato} E.,  2011, \mn@doi [\mnras] {10.1111/j.1365-2966.2010.17449.x}, \href {https://ui.adsabs.harvard.edu/abs/2011MNRAS.410..381B} {410, 381}

\bibitem[\protect\citeauthoryear{{Cao} et~al.,}{{Cao} et~al.}{2019}]{2019arXiv190502773C}
{Cao} Z.,  et~al., 2019, \mn@doi [arXiv e-prints] {10.48550/arXiv.1905.02773}, \href {https://ui.adsabs.harvard.edu/abs/2019arXiv190502773C} {p. arXiv:1905.02773}

\bibitem[\protect\citeauthoryear{{Carilli}}{{Carilli}}{1996}]{1996A&A...305..402C}
{Carilli} C.~L.,  1996, \aap, \href {https://ui.adsabs.harvard.edu/abs/1996A&A...305..402C} {305, 402}

\bibitem[\protect\citeauthoryear{{Cataldo}, {Pagliaroli}, {Vecchiotti}  \& {Villante}}{{Cataldo} et~al.}{2020}]{2020ApJ...904...85C}
{Cataldo} M.,  {Pagliaroli} G.,  {Vecchiotti} V.,   {Villante} F.~L.,  2020, \mn@doi [\apj] {10.3847/1538-4357/abc0ee}, \href {https://ui.adsabs.harvard.edu/abs/2020ApJ...904...85C} {904, 85}

\bibitem[\protect\citeauthoryear{{Chen}, {Liu}  \& {Wang}}{{Chen} et~al.}{2023}]{Chen:2023uK}
{Chen} X.-B.,  {Liu} R.-Y.,   {Wang} X.-Y.,  2023, \mn@doi [PoS] {10.22323/1.444.0815}, ICRC2023, 815

\bibitem[\protect\citeauthoryear{{Chevalier}}{{Chevalier}}{2000}]{2000ApJ...539L..45C}
{Chevalier} R.~A.,  2000, \mn@doi [\apjl] {10.1086/312835}, \href {https://ui.adsabs.harvard.edu/abs/2000ApJ...539L..45C} {539, L45}

\bibitem[\protect\citeauthoryear{{Draine}}{{Draine}}{2011}]{2011piim.book.....D}
{Draine} B.~T.,  2011, {Physics of the Interstellar and Intergalactic Medium}.
Princeton University Press

\bibitem[\protect\citeauthoryear{{Faucher-Gigu{\`e}re} \& {Kaspi}}{{Faucher-Gigu{\`e}re} \& {Kaspi}}{2006}]{2006ApJ...643..332F}
{Faucher-Gigu{\`e}re} C.-A.,  {Kaspi} V.~M.,  2006, \mn@doi [\apj] {10.1086/501516}, \href {https://ui.adsabs.harvard.edu/abs/2006ApJ...643..332F} {643, 332}

\bibitem[\protect\citeauthoryear{Fiori, Olmi, Amato, Bandiera, Bucciantini, Zampieri  \& Burtovoi}{Fiori et~al.}{2022}]{Fiori_2022}
Fiori M.,  Olmi B.,  Amato E.,  Bandiera R.,  Bucciantini N.,  Zampieri L.,   Burtovoi A.,  2022, \mn@doi [Monthly Notices of the Royal Astronomical Society] {10.1093/mnras/stac019}, 511, 1439–1453

\bibitem[\protect\citeauthoryear{{Gaensler} \& {Slane}}{{Gaensler} \& {Slane}}{2006}]{2006ARA&A..44...17G}
{Gaensler} B.~M.,  {Slane} P.~O.,  2006, \mn@doi [\araa] {10.1146/annurev.astro.44.051905.092528}, \href {https://ui.adsabs.harvard.edu/abs/2006ARA&A..44...17G} {44, 17}

\bibitem[\protect\citeauthoryear{{Gelfand}, {Slane}  \& {Zhang}}{{Gelfand} et~al.}{2009}]{2009ApJ...703.2051G}
{Gelfand} J.~D.,  {Slane} P.~O.,   {Zhang} W.,  2009, \mn@doi [\apj] {10.1088/0004-637X/703/2/2051}, \href {https://ui.adsabs.harvard.edu/abs/2009ApJ...703.2051G} {703, 2051}

\bibitem[\protect\citeauthoryear{{H.E.S.S. Collaboration} et~al.,}{{H.E.S.S. Collaboration} et~al.}{2018a}]{2018A&A...612A...2H}
{H.E.S.S. Collaboration} et~al., 2018a, \mn@doi [\aap] {10.1051/0004-6361/201629377}, \href {https://ui.adsabs.harvard.edu/abs/2018A&A...612A...2H} {612, A2}

\bibitem[\protect\citeauthoryear{{H.E.S.S. Collaboration} et~al.,}{{H.E.S.S. Collaboration} et~al.}{2018b}]{2018A&A...617A..73H}
{H.E.S.S. Collaboration} et~al., 2018b, \mn@doi [\aap] {10.1051/0004-6361/201833202}, \href {https://ui.adsabs.harvard.edu/abs/2018A&A...617A..73H} {617, A73}

\bibitem[\protect\citeauthoryear{{Ha}, {Ryu}  \& {Kang}}{{Ha} et~al.}{2021}]{2021ApJ...907...26H}
{Ha} J.-H.,  {Ryu} D.,   {Kang} H.,  2021, \mn@doi [\apj] {10.3847/1538-4357/abd247}, \href {https://ui.adsabs.harvard.edu/abs/2021ApJ...907...26H} {907, 26}

\bibitem[\protect\citeauthoryear{{Hobbs}, {Lorimer}, {Lyne}  \& {Kramer}}{{Hobbs} et~al.}{2005}]{2005MNRAS.360..974H}
{Hobbs} G.,  {Lorimer} D.~R.,  {Lyne} A.~G.,   {Kramer} M.,  2005, \mn@doi [\mnras] {10.1111/j.1365-2966.2005.09087.x}, \href {https://ui.adsabs.harvard.edu/abs/2005MNRAS.360..974H} {360, 974}

\bibitem[\protect\citeauthoryear{{Johnston}, {Smith}, {Karastergiou}  \& {Kramer}}{{Johnston} et~al.}{2020}]{2020MNRAS.497.1957J}
{Johnston} S.,  {Smith} D.~A.,  {Karastergiou} A.,   {Kramer} M.,  2020, \mn@doi [\mnras] {10.1093/mnras/staa2110}, \href {https://ui.adsabs.harvard.edu/abs/2020MNRAS.497.1957J} {497, 1957}

\bibitem[\protect\citeauthoryear{{Kargaltsev} \& {Pavlov}}{{Kargaltsev} \& {Pavlov}}{2010}]{2010AIPC.1248...25K}
{Kargaltsev} O.,  {Pavlov} G.~G.,  2010, in {Comastri} A.,  {Angelini} L.,   {Cappi} M.,  eds,  American Institute of Physics Conference Series Vol. 1248, X-ray Astronomy 2009; Present Status, Multi-Wavelength Approach and Future Perspectives. pp 25--28 (\mn@eprint {arXiv} {1002.0885}), \mn@doi{10.1063/1.3475228}

\bibitem[\protect\citeauthoryear{{Krumholz}, {Crocker}, {Xu}, {Lazarian}, {Rosevear}  \& {Bedwell-Wilson}}{{Krumholz} et~al.}{2020}]{2020MNRAS.493.2817K}
{Krumholz} M.~R.,  {Crocker} R.~M.,  {Xu} S.,  {Lazarian} A.,  {Rosevear} M.~T.,   {Bedwell-Wilson} J.,  2020, \mn@doi [\mnras] {10.1093/mnras/staa493}, \href {https://ui.adsabs.harvard.edu/abs/2020MNRAS.493.2817K} {493, 2817}

\bibitem[\protect\citeauthoryear{{Lacki}, {Thompson}, {Quataert}, {Loeb}  \& {Waxman}}{{Lacki} et~al.}{2011}]{2011ApJ...734..107L}
{Lacki} B.~C.,  {Thompson} T.~A.,  {Quataert} E.,  {Loeb} A.,   {Waxman} E.,  2011, \mn@doi [\apj] {10.1088/0004-637X/734/2/107}, \href {https://ui.adsabs.harvard.edu/abs/2011ApJ...734..107L} {734, 107}

\bibitem[\protect\citeauthoryear{{Mannheim}, {Els{\"a}sser}  \& {Tibolla}}{{Mannheim} et~al.}{2012}]{2012APh....35..797M}
{Mannheim} K.,  {Els{\"a}sser} D.,   {Tibolla} O.,  2012, \mn@doi [Astroparticle Physics] {10.1016/j.astropartphys.2012.02.009}, \href {https://ui.adsabs.harvard.edu/abs/2012APh....35..797M} {35, 797}

\bibitem[\protect\citeauthoryear{{Martin} \& {Torres}}{{Martin} \& {Torres}}{2022}]{2022JHEAp..36..128M}
{Martin} J.,  {Torres} D.~F.,  2022, \mn@doi [Journal of High Energy Astrophysics] {10.1016/j.jheap.2022.09.003}, \href {https://ui.adsabs.harvard.edu/abs/2022JHEAp..36..128M} {36, 128}

\bibitem[\protect\citeauthoryear{{Martin}, {Tibaldo}, {Marcowith}  \& {Abdollahi}}{{Martin} et~al.}{2022}]{2022A&A...666A...7M}
{Martin} P.,  {Tibaldo} L.,  {Marcowith} A.,   {Abdollahi} S.,  2022, \mn@doi [\aap] {10.1051/0004-6361/202244002}, \href {https://ui.adsabs.harvard.edu/abs/2022A&A...666A...7M} {666, A7}

\bibitem[\protect\citeauthoryear{{Mayer}, {Brucker}, {Holler}, {Jung}, {Valerius}  \& {Stegmann}}{{Mayer} et~al.}{2012}]{2012arXiv1202.1455M}
{Mayer} M.,  {Brucker} J.,  {Holler} M.,  {Jung} I.,  {Valerius} K.,   {Stegmann} C.,  2012, \mn@doi [arXiv e-prints] {10.48550/arXiv.1202.1455}, \href {https://ui.adsabs.harvard.edu/abs/2012arXiv1202.1455M} {p. arXiv:1202.1455}

\bibitem[\protect\citeauthoryear{{Melo}, {P{\'e}rez Garc{\'\i}a}, {Acosta-Pulido}, {Mu{\~n}oz-Tu{\~n}{\'o}n}  \& {Rodr{\'\i}guez Espinosa}}{{Melo} et~al.}{2002}]{2002ApJ...574..709M}
{Melo} V.~P.,  {P{\'e}rez Garc{\'\i}a} A.~M.,  {Acosta-Pulido} J.~A.,  {Mu{\~n}oz-Tu{\~n}{\'o}n} C.,   {Rodr{\'\i}guez Espinosa} J.~M.,  2002, \mn@doi [\apj] {10.1086/341109}, \href {https://ui.adsabs.harvard.edu/abs/2002ApJ...574..709M} {574, 709}

\bibitem[\protect\citeauthoryear{{Ohm} \& {Hinton}}{{Ohm} \& {Hinton}}{2013}]{2013MNRAS.429L..70O}
{Ohm} S.,  {Hinton} J.~A.,  2013, \mn@doi [\mnras] {10.1093/mnrasl/sls025}, \href {https://ui.adsabs.harvard.edu/abs/2013MNRAS.429L..70O} {429, L70}

\bibitem[\protect\citeauthoryear{{Peretti}, {Blasi}, {Aharonian}  \& {Morlino}}{{Peretti} et~al.}{2019}]{2019MNRAS.487..168P}
{Peretti} E.,  {Blasi} P.,  {Aharonian} F.,   {Morlino} G.,  2019, \mn@doi [\mnras] {10.1093/mnras/stz1161}, \href {https://ui.adsabs.harvard.edu/abs/2019MNRAS.487..168P} {487, 168}

\bibitem[\protect\citeauthoryear{{Persic}, {Rephaeli}  \& {Arieli}}{{Persic} et~al.}{2008}]{2008A&A...486..143P}
{Persic} M.,  {Rephaeli} Y.,   {Arieli} Y.,  2008, \mn@doi [\aap] {10.1051/0004-6361:200809525}, \href {https://ui.adsabs.harvard.edu/abs/2008A&A...486..143P} {486, 143}

\bibitem[\protect\citeauthoryear{{Pietsch} et~al.,}{{Pietsch} et~al.}{2001}]{2001A&A...365L.174P}
{Pietsch} W.,  et~al., 2001, \mn@doi [\aap] {10.1051/0004-6361:20000068}, \href {https://ui.adsabs.harvard.edu/abs/2001A&A...365L.174P} {365, L174}

\bibitem[\protect\citeauthoryear{{Porter} \& {Strong}}{{Porter} \& {Strong}}{2005}]{2005ICRC....4...77P}
{Porter} T.~A.,  {Strong} A.~W.,  2005, in 29th International Cosmic Ray Conference (ICRC29), Volume 4. p.~77 (\mn@eprint {arXiv} {astro-ph/0507119}), \mn@doi{10.48550/arXiv.astro-ph/0507119}

\bibitem[\protect\citeauthoryear{{Rekola}, {Richer}, {McCall}, {Valtonen}, {Kotilainen}  \& {Flynn}}{{Rekola} et~al.}{2005}]{2005MNRAS.361..330R}
{Rekola} R.,  {Richer} M.~G.,  {McCall} M.~L.,  {Valtonen} M.~J.,  {Kotilainen} J.~K.,   {Flynn} C.,  2005, \mn@doi [\mnras] {10.1111/j.1365-2966.2005.09166.x}, \href {https://ui.adsabs.harvard.edu/abs/2005MNRAS.361..330R} {361, 330}

\bibitem[\protect\citeauthoryear{{Shimono}, {Totani}  \& {Sudoh}}{{Shimono} et~al.}{2021}]{2021MNRAS.506.6212S}
{Shimono} N.,  {Totani} T.,   {Sudoh} T.,  2021, \mn@doi [\mnras] {10.1093/mnras/stab2118}, \href {https://ui.adsabs.harvard.edu/abs/2021MNRAS.506.6212S} {506, 6212}

\bibitem[\protect\citeauthoryear{{Tanaka} \& {Kashiyama}}{{Tanaka} \& {Kashiyama}}{2023}]{2023MNRAS.525.2750T}
{Tanaka} S.~J.,  {Kashiyama} K.,  2023, \mn@doi [\mnras] {10.1093/mnras/stad2504}, \href {https://ui.adsabs.harvard.edu/abs/2023MNRAS.525.2750T} {525, 2750}

\bibitem[\protect\citeauthoryear{{Thompson}, {Quataert}, {Waxman}, {Murray}  \& {Martin}}{{Thompson} et~al.}{2006}]{2006ApJ...645..186T}
{Thompson} T.~A.,  {Quataert} E.,  {Waxman} E.,  {Murray} N.,   {Martin} C.~L.,  2006, \mn@doi [\apj] {10.1086/504035}, \href {https://ui.adsabs.harvard.edu/abs/2006ApJ...645..186T} {645, 186}

\bibitem[\protect\citeauthoryear{{Torres}, {Cillis}, {Mart{\'\i}n}  \& {de O{\~n}a Wilhelmi}}{{Torres} et~al.}{2014}]{2014JHEAp...1...31T}
{Torres} D.~F.,  {Cillis} A.,  {Mart{\'\i}n} J.,   {de O{\~n}a Wilhelmi} E.,  2014, \mn@doi [Journal of High Energy Astrophysics] {10.1016/j.jheap.2014.02.001}, \href {https://ui.adsabs.harvard.edu/abs/2014JHEAp...1...31T} {1, 31}

\bibitem[\protect\citeauthoryear{{VERITAS Collaboration} et~al.,}{{VERITAS Collaboration} et~al.}{2009}]{2009Natur.462..770V}
{VERITAS Collaboration} et~al., 2009, \mn@doi [\nat] {10.1038/nature08557}, \href {https://ui.adsabs.harvard.edu/abs/2009Natur.462..770V} {462, 770}

\bibitem[\protect\citeauthoryear{{Wang} \& {Fields}}{{Wang} \& {Fields}}{2018}]{2018MNRAS.474.4073W}
{Wang} X.,  {Fields} B.~D.,  2018, \mn@doi [\mnras] {10.1093/mnras/stx2917}, \href {https://ui.adsabs.harvard.edu/abs/2018MNRAS.474.4073W} {474, 4073}

\bibitem[\protect\citeauthoryear{{Watters} \& {Romani}}{{Watters} \& {Romani}}{2011}]{2011ApJ...727..123W}
{Watters} K.~P.,  {Romani} R.~W.,  2011, \mn@doi [\apj] {10.1088/0004-637X/727/2/123}, \href {https://ui.adsabs.harvard.edu/abs/2011ApJ...727..123W} {727, 123}

\bibitem[\protect\citeauthoryear{{Wik} et~al.,}{{Wik} et~al.}{2014}]{2014ApJ...797...79W}
{Wik} D.~R.,  et~al., 2014, \mn@doi [\apj] {10.1088/0004-637X/797/2/79}, \href {https://ui.adsabs.harvard.edu/abs/2014ApJ...797...79W} {797, 79}

\bibitem[\protect\citeauthoryear{{Yan} \& {Liu}}{{Yan} \& {Liu}}{2023}]{2023PhRvD.107j3028Y}
{Yan} K.,  {Liu} R.-Y.,  2023, \mn@doi [\prd] {10.1103/PhysRevD.107.103028}, \href {https://ui.adsabs.harvard.edu/abs/2023PhRvD.107j3028Y} {107, 103028}

\bibitem[\protect\citeauthoryear{{Zhang}, {Chen}  \& {Fang}}{{Zhang} et~al.}{2008}]{2008ApJ...676.1210Z}
{Zhang} L.,  {Chen} S.~B.,   {Fang} J.,  2008, \mn@doi [\apj] {10.1086/527466}, \href {https://ui.adsabs.harvard.edu/abs/2008ApJ...676.1210Z} {676, 1210}

\bibitem[\protect\citeauthoryear{{Zhu}, {Zhang}  \& {Fang}}{{Zhu} et~al.}{2018}]{2018A&A...609A.110Z}
{Zhu} B.-T.,  {Zhang} L.,   {Fang} J.,  2018, \mn@doi [\aap] {10.1051/0004-6361/201629108}, \href {https://ui.adsabs.harvard.edu/abs/2018A&A...609A.110Z} {609, A110}

\bibitem[\protect\citeauthoryear{{de O{\~n}a Wilhelmi}, {L{\'o}pez-Coto}, {Amato}  \& {Aharonian}}{{de O{\~n}a Wilhelmi} et~al.}{2022}]{2022ApJ...930L...2D}
{de O{\~n}a Wilhelmi} E.,  {L{\'o}pez-Coto} R.,  {Amato} E.,   {Aharonian} F.,  2022, \mn@doi [\apjl] {10.3847/2041-8213/ac66cf}, \href {https://ui.adsabs.harvard.edu/abs/2022ApJ...930L...2D} {930, L2}

\makeatother
\end{thebibliography}
\bsp	
\label{lastpage}
\end{document}